\documentclass[aps,prb,twocolumn,showpacs,superscriptaddress,longbibliography]{revtex4-2}
\usepackage{amsmath,amssymb,graphicx}
\usepackage{graphicx,epstopdf}
\usepackage{gensymb}
\usepackage[dvipsnames]{xcolor}
\usepackage[T1]{fontenc}
\usepackage[utf8]{inputenc}
\newcommand{\be}{\begin{equation}}
	\newcommand{\ee}{\end{equation}}
\newcommand{\bea}{\begin{eqnarray}}
	\newcommand{\eea}{\end{eqnarray}}
\newcommand{\bse}{\begin{subequations}}
	\newcommand{\ese}{\end{subequations}}

\usepackage{multibib}
\usepackage{color}
\usepackage[colorlinks,bookmarks=false,citecolor=darkblue,linkcolor=red,urlcolor=blue]{hyperref}

\definecolor{darkred}{rgb}{0.7,0.0,0.0}

\definecolor{darkblue}{rgb}{0,0.02,0.45}

\definecolor{darkgreen}{rgb}{0.02,0.45,0.0}

\definecolor{violet}{rgb}{0.8,0.2,0.6}

\usepackage{float}
\usepackage{natbib}
\usepackage{mathrsfs}
\usepackage{soul}
\setstcolor{red}

\definecolor{darkgreen}{rgb}{0, 0.4, 0}

\begin{document}
\title{Large magnetocaloric effect in the kagome ferromagnet Li$_9$Cr$_3$(P$_2$O$_7$)$_3$(PO$_4$)$_2$}
\author{Akshata Magar}
\author{Somesh K}
\author{Vikram Singh}
\affiliation{School of Physics, Indian Institute of Science Education and Research Thiruvananthapuram-695551, India}
\author{J.~J.~Abraham}
\affiliation{Leibniz IFW Dresden, D-01069 Dresden, Germany}
\affiliation{Institute for Solid State and Materials Physics, TU Dresden, 01069 Dresden, Germany}
\author{Y.~Senyk}
\author{A.~Alfonsov}
\affiliation{Leibniz IFW Dresden, D-01069 Dresden, Germany}
\author{B.~B\"uchner}
\affiliation{Leibniz IFW Dresden, D-01069 Dresden, Germany}
\affiliation{Institute for Solid State and Materials Physics and W{\"u}rzburg-Dresden Cluster of Excellence ct.qmat, TU Dresden, D-01062 Dresden, Germany}
\author{V.~Kataev}
\affiliation{Leibniz IFW Dresden, D-01069 Dresden, Germany}
\author{A. A. Tsirlin}
\affiliation{Felix Bloch Institute for Solid-State Physics, Leipzig University, 04103 Leipzig, Germany}
\author{R. Nath}
\email{rnath@iisertvm.ac.in}
\affiliation{School of Physics, Indian Institute of Science Education and Research Thiruvananthapuram-695551, India}

\begin{abstract}
Single-crystal growth, magnetic properties, and magnetocaloric effect of the $S = 3/2$ kagome ferromagnet Li$_9$Cr$_3$(P$_2$O$_7$)$_3$(PO$_4$)$_2$ (trigonal, space group: $P\bar{3}c1$) are reported. Magnetization data suggest dominant ferromagnetic intra-plane coupling with a weak anisotropy and the onset of ferromagnetic ordering at $T_{\rm C} \simeq 2.6$~K. Microscopic analysis reveals a very small ratio of interlayer to intralayer ferromagnetic couplings ($J_{\perp}/J \simeq 0.02$). Electron spin resonance data suggest the presence of short-range correlations above $T_{\rm C}$ and confirms quasi-two-dimensional character of the spin system. A large magnetocaloric effect characterized by isothermal entropy change of $-\Delta S_{\rm m}\simeq 31$~J~kg$^{-1}$~K$^{-1}$ and adiabatic temperature change of $-\Delta T_{\rm ad}\simeq 9$~K upon a field sweep of 7~T is observed around $T_{\rm C}$. This leads to a large relative cooling power of $RCP \simeq 284$~J~kg$^{-1}$. The large magnetocaloric effect, together with negligible hysteresis render Li$_9$Cr$_3$(P$_2$O$_7$)$_3$(PO$_4$)$_2$ a promising material for magnetic refrigeration at low temperatures. The magnetocrystalline anisotropy constant $K \simeq -7.42 \times 10^4$~erg~cm$^{-3}$ implies that the compound is an easy-plane type ferromagnet with the hard axis normal to the $ab$-plane, consistent with the magnetization data.
\end{abstract}
\maketitle

\section{Introduction}

Kagome lattice hosts a plethora of interesting phenomena. Its frustrated nature renders antiferromagnetic kagome insulators a natural playground for the experimental realization of quantum spin liquid~\cite{Mendels455,*Han406,*Carrasquilla7421}. Whereas kagome ferromagnets are not frustrated and develop magnetic order, they are no less interesting because flat bands and Dirac fermions expected in this setting have far-reaching implications for transport properties. Recent work on ferromagnetic kagome metals exposed anomalous Hall and Nernst effects, as well as chiral edge states, in several intermetallic compounds, such as Fe$_3$Sn$_2$~\cite{Linda638}, Co$_3$Sn$_2$S$_2$~\cite{Howard4269}, LiMn$_6$Sn$_6$~\cite{Chen144410}, and UCo$_{0.8}$Ru$_{0.2}$Al~\cite{Asabaeabf1467}. Concurrently, insulating kagome ferromagnets were actively studied in the context of magnon Hall effect and other exotic properties associated with Dirac magnons~\cite{Mook134409,Chisnell147201,Chisnell214403}. 


Ferromagnets are further interesting as materials with the large magnetocaloric effect (MCE) that can be instrumental in cooling via adiabatic demagnetization for reaching temperatures in the sub-Kelvin range~\cite{Gschneidner1479,Pecharsky44,Kitanovski201903741}. This magnetic refrigeration technique is often considered as the most energy-efficient, cost-effective (as $^3$He and $^4$He are expensive), and environment-friendly replacement for the conventional refrigeration based on gas compression/expansion technique. For this purpose, materials with large magnetic moment, low magnetic anisotropy, low magnetic hysteresis, and extremely low transition temperature are desirable~\cite{Franco112,Phan325}. The nature of the magnetic transition and the specific form of the magnetic structure are also deciding factors for the performance of a MCE material.  Ferromagnetic insulators with second-order phase transition are proposed to be excellent MCE materials, as only a small change in applied magnetic field is sufficient to yield a large entropy change and adiabatic temperature change, compared to any paramagnetic salt~\cite{Pecharsky44,Phan325}.
A very few ferromagnetic insulators with low transition temperature are reported to satisfy the above prerequisites and qualify for low-temperature applications~\cite{Krishna355001,Roy064412,Lingwei152403}.

In the following, we report the magnetic properties of Li$_9$Cr$_3$(P$_2$O$_7$)$_3$(PO$_4$)$_2$ (LCPP) which is a structural sibling of the recently reported $S=5/2$ Heisenberg kagome antiferromagnet  Li$_9$Fe$_3$(P$_2$O$_7$)$_3$(PO$_4$)$_2$ (LFPP) with a trigonal space group $P\bar{3}c1$~\cite{Kermarrec157202}. LFPP shows the onset of an antiferromagnetic (AFM) ordering below $T_{\rm N} \simeq 1.3$~K and a characteristic $1/3$ magnetization plateau below $T^* \simeq 5$~K. The NMR spectra along with the NMR spin-lattice relaxation time reveal the presence of an exotic semiclassical nematic spin liquid regime between $T_{\rm N}$ and $T^*$.
On the contrary, LCPP is found to be a ferromagnet and it undergoes a ferromagnetic (FM) ordering at $T_{\rm C} \simeq 2.5$~K. LCPP exhibits a large MCE around $T_{\rm C}$ and appear to have strong potential for cryogenic applications such as low temperature sensors in space research, achieving sub-kelvin temperatures for basic research, hydrogen and helium gas liquefaction etc~\cite{Gschneidner1479,Martinez4301}.

\begin{figure}
	\includegraphics[width=\columnwidth] {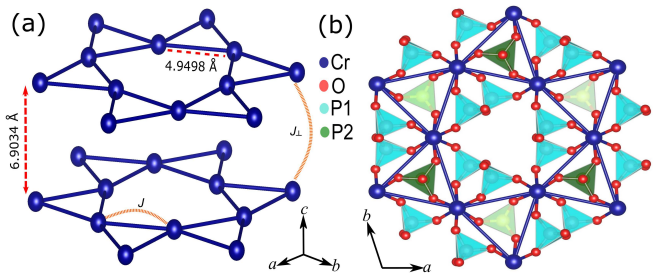}	\caption{\label{Fig1}(a) Corner-sharing equilateral triangles of Cr$^{3+}$ form a kagome lattice. The intraplane ($J$) and interplanar ($J_{\perp}$) coupling are shown. The kagome lattice layers are well separated from each other with interlayer distance of 6.9034~\AA. (b) Interactions between the magnetic Cr$^{3+}$ ions via PO$_4$ tetrahedra are shown.}
\end{figure}
\section{Methods}
Platelet single crystals of LCPP with the lateral size of 0.5~mm to 1~mm were synthesized by a self-flux technique as reported in Ref.~\cite{Poisson32}. The mixture of starting materials, Li$_3$PO$_4$, Cr$_2$O$_3$, and NH$_4$H$_2$PO$_4$ in the molar ratio 15:1:9 was kept in an alumina crucible and heated gradually to $900\degree$C. The cooling process involves three steps. At first, the sample was cooled down to $850\degree$C at a rate of $50\degree$C per hour and then to $600\degree$C at a slow rate of $2\degree$C per hour. Finally, the sample was allowed to cool naturally to room temperature. In order to dissolve the flux and separate the crystals, the sample was treated with 1~M solution of acetic acid for five days followed by the treatment with saturated NaCl solution and distilled water. The final product after the treatment yields the mixture of mm-sized single crystals and polycrystalline sample. The large-sized crystals were hand-picked and the remaining part is grinded to get the polycrystalline sample.

Room-temperature single-crystal x-ray diffraction (XRD) was performed on a good quality single crystal using the Bruker KAPPA APEX-II CCD diffractometer equipped with graphite monochromated Mo $K_{\alpha 1}$ radiation ($\lambda = 0.71073$~\AA). The APEX3 software was used to collect the data that were further reduced with SAINT/XPREP followed by an empirical absorption correction using the SADABS program~\cite{Sheldrick1994}. The phase purity of the polycrystalline sample was confirmed from powder XRD (PANalytical Xpert-Pro, Cu~$K_\alpha$ radiation with $\lambda_{av}=1.54182$~\AA). The temperature-dependent powder XRD measurement was performed in the temperature range 15~K~$\leq T \leq$~300~K with a low-temperature (Oxford Phenix) attachment to the diffractometer.

Magnetization ($M$) measurement was performed as a function of temperature ($T$) and magnetic field ($H$) using a superconducting quantum interference device (SQUID) (MPMS-3, Quantum Design) magnetometer. The data were collected in the temperature range 1.8~K $\leq T \leq$ 350~K and in the magnetic field range 0 $\leq H \leq$ 7~T.  Heat capacity ($C_{\rm p}$) as a function of $T$ (0.5~K~$\leq T \leq$~300~K) and $H$ was measured on a small piece of sintered pellet using the relaxation technique in the physical property measurement system (PPMS, Quantum Design). Measurements below 2~K were carried out using an additional $^{3}$He insert in the PPMS.

High-field electron spin resonance (HF-ESR) spectroscopy was used to study the single crystals of LCPP. For the measurements in a frequency range 75 - 330~GHz, a vector network analyzer (PNA-X from Keysight Technologies) was used and for frequencies up to 975~GHz a modular Amplifier/Multiplier Chain (AMC from Virginia Diodes Inc.) was used for the generation of microwaves in combination with a hot electron InSb bolometer for detection. All measurements were performed at a given fixed frequency in the field-sweep mode up to 16~T, using a superconducting magnet system from Oxford Inst. The sample was mounted onto a transmission probe head which is then inserted in a $^4$He variable temperature insert (VTI) of the magnet cryostat to enable measurements in a temperature range of $1.8-300$~K.

Density-functional (DFT) band-structure calculations were performed in the \texttt{FPLO} code~\cite{fplo} with the Perdew-Burke-Ernzerhof flavor of the exchange-correlation potential~\cite{pbe96}. Correlation effects in the Cr $3d$ shell were included on the mean-field level within DFT+$U$ using the on-site Coulomb repulsion parameter $U_d=2$\,eV, Hund's coupling $J_d=1$\,eV, and double-counting correction in the atomic limit~\cite{janson2014,Somesh104422}. Exchange couplings $J_{ij}$ were obtained by mapping~\cite{xiang2011} total energies of collinear magnetic configurations onto the spin Hamiltonian,
\begin{equation}
	\mathcal H=\sum_{\langle ij\rangle} J_{ij}\mathbf S_i\mathbf S_j
	\label{eq:ham}\end{equation} 
where the summation is over pairs, and $S=\frac32$. Energies were converged on a $k$ mesh with 64 points within the first Brillouin zone. Thermodynamic properties for the model defined by Eq.~\eqref{eq:ham} were obtained from quantum Monte-Carlo simulations performed with the \texttt{loop} algorithm~\cite{loop} of the \texttt{ALPS} simulation package~\cite{alps}. Finite lattices with up to 752 sites and periodic boundary conditions were used.

\section{Results and Discussion}
\subsection{X-ray Diffraction}
  \begin{table}[h]
	\caption{Crystallographic data for LCPP at room temperature, obtained from single-crystal XRD.}
	\begin{tabular}{ c   c } 
		\hline\hline
		Empirical formula & Cr$_3$ Li$_9$ O$_{29}$ P$_8$ \\ 
		Formula weight(M$_r$)  & 930.22~g~mol$^{-1}$ \\ 
		Temperature & 296(2) K \\ 
		Crystal system  & Trigonal \\
		Space group  & $P\bar{3}c1$ \\
		Lattice parameters   & $a=9.668(3)$~\AA~~~$\alpha = 90\degree$\\& $b=9.668(3)$~\AA~~~$\beta = 90\degree$\\& $c=13.610(6)$~\AA~~~$\gamma = 120\degree$\\	
		Unit cell volume & 1101.7(8)~\AA$^3$\\
		$Z$ & 2\\
		Density (calculated) &	2.804~g~cm$^{-3}$\\
		Wavelength & 0.71073 \AA \\
		Radiation type & MoK$\alpha_1$\\
		Diffractometer & Bruker KAPPA 
		APEX-II CCD\\
		Crystal size & $0.049\times 0.035\times 0.027$~mm$^3$\\
		2$\theta$ range & 2.993 to 25.997$\degree$\\
		Index ranges & $-11\leq h\leq 11$\\& $-11\leq k\leq 11$ \\& $-16\leq l \leq 16$ \\
		$F(000)$ & 902\\
		Reflections collected & 6940\\
		Independent reflections & 735 [$R_{int}$ = 0.0454]\\
		Data / restraints / parameters & 735/0/76\\
		Goodness-of-fit on $F^2$ & 1.098\\
		Final $R$ indices [$I$~$\geq~$2$\sigma$($I$)] &	$R1$ = 0.0304,\\& $\omega$$R2$ = 0.0892\\
		$R$ indices(all data) & 	$R1$ = 0.0351,\\& $\omega$$R2$ = 0.0919\\
		Largest diff. peak and hole & 	$+0.445/-0.970$ e.\AA$^{-3}$\\
		\hline\hline
	\end{tabular}
	\label{Table1}
\end{table}
The crystal structure of LCPP was solved from single-crystal XRD data with direct methods using SHELXT-2018/2~\cite{Sheldrick2015shelxt} and refined by the full matrix least squares on $F^2$ using SHELXL-2018/3, respectively~\cite{Sheldrick2018shelxl}. Details of the crystal structure and the refined parameters are summarized in Table~\ref{Table1}. LCPP crystallizes in the trigonal space group $P\bar{3}c1$ (No.~165). The refined atomic positions at room temperature are listed in Table~\ref{Table2}. These structural parameters are in good agreement with the previous report~\cite{Poisson32}.

The schematic view of the crystal structure of LCPP is presented in Fig.~\ref{Fig1}. It illustrates the corner sharing of CrO$_6$ octahedra and PO$_4$ tetrahedra forming equilateral triangles with a geometrically deformed but regular kagome lattice. Though all of the Cr$^{3+}$ -- Cr$^{3+}$ distances in each hexagon are equal ($\sim 4.949$~\AA), the bond angles are different: three angles are about $\sim 146.3\degree$ and the remaining three angles are about $\sim 93.7\degree$. The nearest-neighbour (NN) coupling between the Cr$^{3+}$ ions in the $ab$-plane is denoted by $J$, while the shortest interplane distance of $\sim 6.903$~\AA~leads to a weak coupling ($J_{\perp}$) between the planes.

\begin{table}
	\caption{Crystal structure of LCPP refined using single-crystal XRD data. The atomic coordinates ($\times 10^4$) and the isotropic atomic displacement parameter $U_{\rm iso}$(\AA$^2\times 10^3$), which is defined as one-third of the trace of the orthogonalized $U_{ij}$ tensor.}
	\begin{tabular}{ c c c c c c }
		\hline\hline
		\multicolumn{1}{p{1cm}}{\centering Atomic \\ sites} & \multicolumn{1}{p{1.3
				cm}}{\centering Wyckoff\\ positions} &\multicolumn{1}{p{2
				cm}}{\centering $x$} &\multicolumn{1}{p{1.3
				cm}}{\centering $y$} &\multicolumn{1}{p{1.3
				cm}}{\centering $z$}  &  $U_{\rm iso}$\\
		\hline
		Cr(1)& $6f$  & 5676(1) & 0 & 2500 & 6(1)\\\vspace*{0.5mm}
		Li(1)& $2b$ &	0 &	0 &	5000 & 19(3)\\
		Li(2)& $12g$ &	3369(7) & 2367(7) &	4374(4) & 16(1)\\
		Li(3)& $4d$ &	6667 &	3333 &	6178(7) & 15(2)\\
		O(1)& $12g$ &	3758(2)	 & $-1053(2)$ &	3331(1) & 9(1)\\
		O(2)& $6f$ &	2120(3) & 0	 & 2500 &	9(1)\\
		O(3)& $12g$ &	792(2) & $-2535(2)$ &	3440(2) & 9(1)\\
		O(4)& $12g$ &	2299(3) & 38(2)  &	4332(2) &	11(1)\\
		O(5)& $12g$ &	6781(2) & 1893(2) &	3352(2) &	10(1)\\
		O(6)& $4d$ &	6667 &	3333 &	4839(3) &	23(1)\\
		P(1)& $12g$ &	2279(1) & $-894(1)$ &	3440(1) &	6(1)\\
		P(2)& $4d$ &	6667 &	3333 &	3736(1) &	5(1)\\
		\hline\hline
	\end{tabular}
	\label{Table2}
\end{table}

\begin{figure}
	\includegraphics[width=\columnwidth]{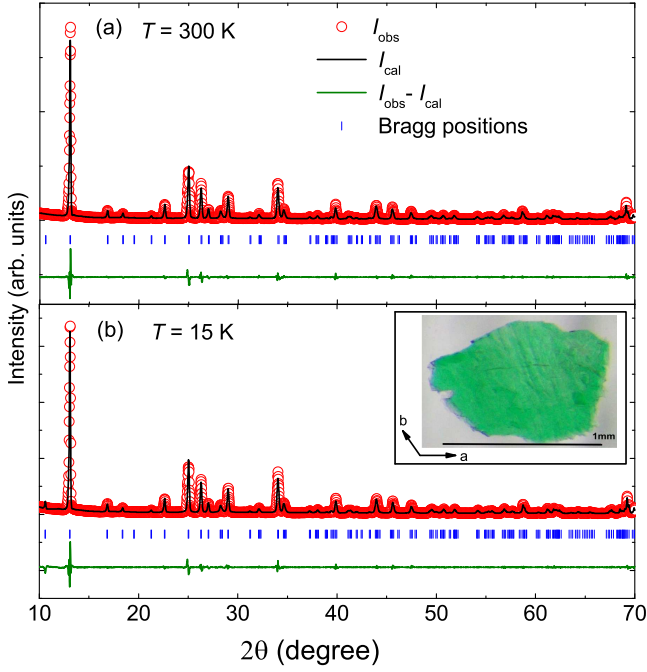}
	\caption{\label{Fig2} Powder XRD data measured at (a) $T = 300$~K and (b) $T = 15$~K. The black solid line represents the Le-Bail fit of the data. Bragg positions are indicated by vertical bars and the solid green line at the bottom denotes the difference between experimental and calculated intensities. The inset of (b) shows a representative single crystal.}
\end{figure}
In order to confirm the phase purity and to scrutinize the presence of any structural distortions, powder XRD data were collected at various temperatures. Le-Bail analysis of the XRD patterns was performed using \verb"FullProf" package~\cite{Juan55} taking the initial structural parameters from the single crystal data (Table~\ref{Table1}). Figure~\ref{Fig2}(a) and (b) present the powder XRD patterns at the highest ($T=300$~K) and lowest ($T=15$~K) measured temperatures, respectively, along with the Le-Beil fits. All the peaks could be indexed based on the space group $P\bar{3}c1$, suggesting phase purity of the polycrystalline sample. The obtained lattice parameters at room temperature are $a = b = 9.6628(3)$~\AA, $c = 13.5769(3)$~\AA, and unit-cell volume $V_{\rm cell} \simeq 1097.85(5)$~\AA$^3$, which are consistent with the single-crystal data. No extra peaks or features were observed in the XRD data corroborating the absence of any structural transition or distortion down to 15~K.

\begin{figure}[h]
	\includegraphics[width=\columnwidth] {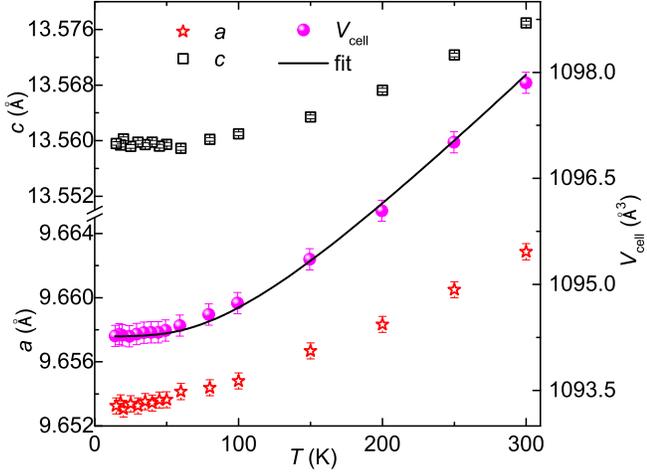}
	\caption{\label{Fig3} Variation of lattice parameters ($a$, $c$, and $V_{\rm cell}$) with temperature. The solid line denotes the fit of $V_{\rm cell}(T)$ by Eq.~\eqref{vcell}.}
\end{figure}
The temperature variation of lattice parameters ($a$, $c$, and $V_{\rm cell}$) is presented in Fig.~\ref{Fig3}. They are found to decrease monotonically upon cooling down to 15~K. $V_{\rm cell}(T)$ was fitted by the equation~\cite{Sebastian104425}
\begin{equation}
V_{\rm cell}(T) = \frac{\gamma U(T)}{K_0} + V_0,
\label{vcell}
\end{equation}
where $V_0$ is the zero-temperature unit cell volume, $K_0$ is the bulk modulus, and $\gamma$ is the Grüneisen parameter. $U(T)$ is the internal energy and it can be expressed in terms of the Debye approximation as
\begin{equation}
 U (T) = 9pk_{B}T(\frac{T}{\theta_{\rm D}})^3 \int_0^{\theta_{\rm D}/T} \frac{x^3}{e^{x}-1} dx.
\end{equation}
Here, $p$ is the number of atoms in the unit cell, $k_{\rm B}$ is the Boltzmann constant, and the Debye temperature is represented by $\theta_{\rm D}$. The fit (see Fig.~\ref{Fig3}) returns $\theta_{\rm D} \simeq 385 $~K, $\gamma/K_0 \simeq 2.06 \times 10^{-4}$~Pa$^{-1}$, and $V_0 \simeq 1094.2$~\AA$^3$. 

\subsection{Magnetization}
\begin{figure}
	\centering
	\includegraphics[width=\columnwidth]{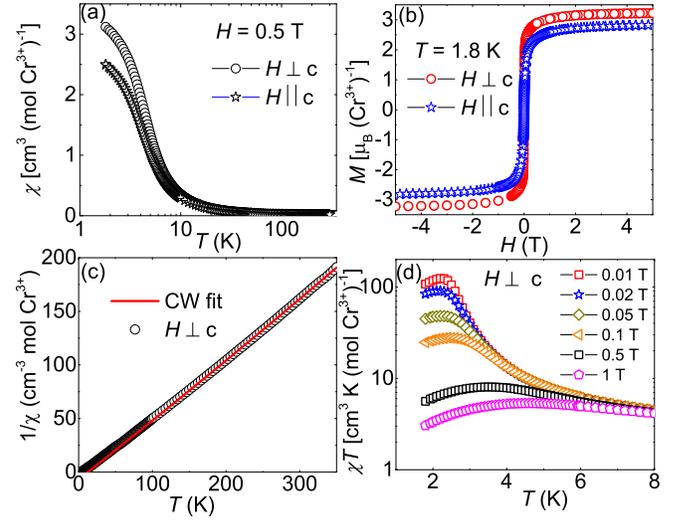}
	\caption{\label{Fig4}(a) $\chi(T)$ measured in the field of $H = 0.5$~T applied perpendicular ($H \perp c$) and parallel ($H\parallel c$) to the $c$-axis. (b) Magnetization at $T = 1.8$~K as a function of applied field for both $H \perp c$ and $H\parallel c$ after correcting for the demagnetization effect. (c) Inverse susceptibility ($1/\chi$) as a function of temperature for $H \perp c$ and solid line is the CW fit. (d) $\chi T$ vs $T$ in different fields for $H \perp c$ in low temperatures.}
\end{figure}
Magnetic susceptibility $\chi(T)~[\equiv M(T)/H]$ measured in an applied field of $H = 0.5$~T perpendicular ($H \perp c$) and parallel ($H \parallel c$) to the kagome plane is displayed in Fig.~\ref{Fig4}(a). With decreasing temperature, $\chi(T)$ increases in a Curie-Weiss manner as expected in the high-temperature paramagnetic (PM) regime, followed by a rapid enhancement at low temperatures. This rapid increase suggests strong FM correlations below about 10~K. Further, the susceptibility for $H \perp c$ and $H \parallel c$ show only a small difference even at low temperatures, which is an indication of weak magnetic anisotropy in the compound.

For a quantitative analysis, we have plotted the inverse susceptibility ($1/\chi$) as a function of temperature in Fig.~\ref{Fig4}(c) for $H \perp c$. For $T \geq 50~\rm K$, it exhibits a completely linear behaviour which was fitted by the modified Curie-Weiss (CW) law
\begin{equation}
\chi(T)=\chi_0+\frac{C}{(T-\theta_{\rm CW})}.
\end{equation} 
Here, $\chi_0$ is the temperature-independent susceptibility, $C$ is the Curie constant, and $\theta_{\rm CW}$ is the CW temperature. The fit yields $\chi_0 \simeq -3.6\times 10^{-4}$~cm$^3$~mol$^{-1}$, $C \simeq 1.92$~cm$^3$~K~mol$^{-1}$, and $\theta_{\rm CW} \simeq 6$~K. From the value of $C$ the effective moment is calculated to be $\mu_{\rm eff} \simeq 3.92~\mu_{\rm B}$ in agreement with the spin-only value of 3.87~$\mu_{\rm B}$ for spin-$3/2$. The positive value of $\theta_{\rm CW}$ suggests that the dominant exchange interactions between Cr$^{3+}$ ions are FM in nature. Using the mean-field expression $J/k_{\rm B} = -3|\theta_{\rm CW}|/z S(S+1)$ with $z=4$ neighbors on the kagome lattice, we estimate $J/k_{\rm B} \simeq -1.2$~K. Moreover, the small peak in $\chi T$ in low magnetic fields reveals $T_{\rm C}\simeq 2.6$~K [Fig.~\ref{Fig4}(d)]. $\chi(T)$ measured in zero-field cooled and field cooled conditions (not shown) in a small magnetic field of $H = 0.01$~T for $H \perp c$ shows no difference, suggesting negligible hysteresis.

The magnetic isotherm ($M$ vs $H$) at $T = 1.8$~K saturates in low fields of $H_{\rm sat}{\perp c} \simeq 0.15$~T and $ H_{\rm sat}{\parallel c}\simeq 0.4$~T with the saturation magnetization of $M_{\rm sat}\sim 3.2~\mu_{\rm B}/{\rm Cr}^{3+}$ and $2.8~\mu_{\rm B}/{\rm Cr}^{3+}$ for $H \perp c$ and $H \parallel c$, respectively (not shown). The obtained $M_{\rm sat}$ values are close to the calculated $M_{\rm sat} = gS\mu_{\rm B} \simeq 2.952\mu_{\rm B}$ and $2.937\mu_{\rm B}$, taking the ESR values $g \simeq 1.968$ and 1.958 for $H \perp c$ and $H \parallel c$, respectively. No visible hysteresis is observed in any of the field directions. A slight difference in the saturation field for $H \perp c$ and $H \parallel c$ may be attributed to the anisotropic demagnetization field caused by the flat shape of the crystals~\cite{Chisnell214403}. The demagnetization factor is negligible when magnetic field is parallel to the crystal plates ($H \perp c$). However, when field is perpendicular to crystal plate ($H \parallel c$), the demagnetization effect is considerably amplified~\cite{Morosan104505}. The data in Fig.~\ref{Fig4}(b) have been corrected for this demagnetizing field as $H_{\rm eff} = H_0 - 4\pi N M$ where $H_{\rm eff}$ and $H_0$ are the effective and applied magnetic fields, respectively, $N$ is the demagnetization factor, and $M$ is magnetic moment in~emu~cm$^{-3}$. To calculate the demagnetization factor in the case of LCPP, we approximated the shape of the sample to a rectangular strip and did the calculation following Ref.~\cite{Prozorov014030}, which yields $N = 0.84$ for $H \parallel c$. The $H_{\rm sat}$ in both directions after demagnetization correction is found to be almost same ($\sim 0.15$~T).

\subsection{Heat Capacity}
\begin{figure}
	\begin{center}
		\includegraphics[width=\columnwidth]{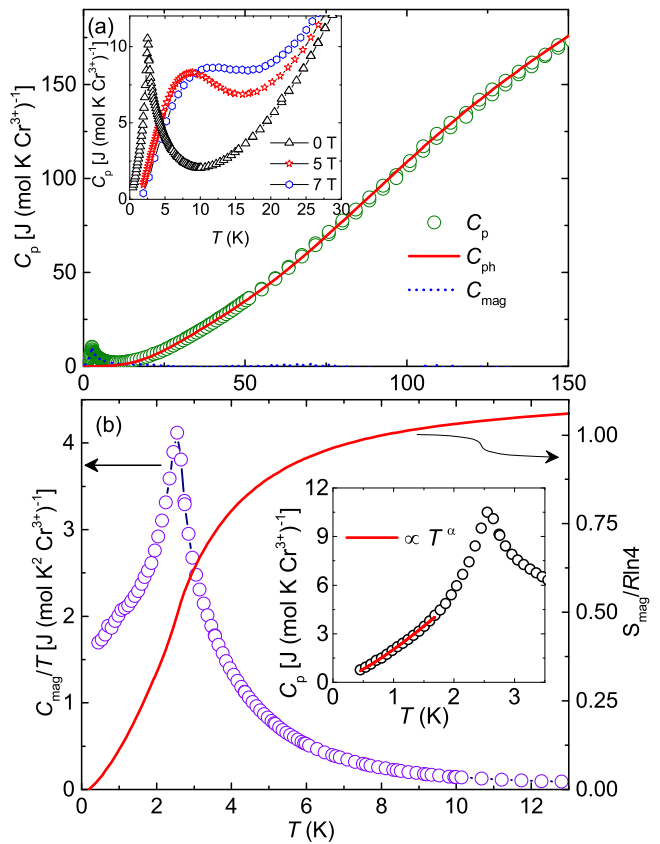} 
		\caption{(a) Heat capacity ($C_{\rm p}$) of LCPP measured in zero applied field. The solid line represents the simulated phonon contribution [$C_{\rm ph}(T)$] and the dotted line represents the magnetic contribution [$C_{\rm mag}(T)$]. Inset: Low temperature $C_{\rm p}$ measured in various applied fields. (b) $C_{\rm mag}/\rm T$ and $S_{\rm mag}/R\ln4$ in left and right y-axes, respectively, are plotted as a function of temperature. Inset: $C_{\rm mag}(T)$ vs $T$. Solid line is the power-law ($C_{\rm mag} = aT^{\alpha}$) fit in the low-$T$ regime.}
		\label{Fig5}
	\end{center}
\end{figure}
Temperature-dependent heat capacity $C_{\rm p}$ measured on the polycrystalline sample is shown in Fig.~\ref{Fig5}.
The $C_{\rm p}$ data exhibit a sharp $\lambda$-type anomaly at $T_{\rm C} \simeq 2.5$~K demonstrating the transition to the magnetically ordered state. Typically, in magnetic insulators, the major contributions to $C_{\rm p}$ are from magnetic ($C_{\rm mag}$) and phonon ($C_{\rm ph}$) parts. In high temperatures, $C_{\rm ph}$ dominates over $C_{\rm mag}$, while at low temperatures it is reverse. One can estimate $C_{\rm mag}$ by subtracting $C_{\rm ph}$ from the total heat capacity. First, we approximate the phonon contribution by fitting the high-$T$ data by a linear combination of one Debye and three Einstein terms as~\cite{Sebastian064413}
\begin{equation}
C_{\rm ph}(T)=f_{\rm D}C_{\rm D}(\theta_{\rm D},T)+\sum_{i = 1}^{3}g_{i}C_{{\rm E}_i}(\theta_{{\rm E}_i},T).
\label{Eq5}
\end{equation}
The first term in Eq.~\eqref{Eq5} is the Debye contribution to $C_{\rm ph}$, which can be written as 
\begin{equation}
C_{\rm D} (\theta_{\rm D}, T)=9nR\left(\frac{T}{\theta_{\rm D}}\right)^{3} \int_0^{\frac{\theta_{\rm D}}{T}}\frac{x^4e^x}{(e^x-1)^2} dx.
\label{Eq6}
\end{equation}
Here, $R$ is the universal gas constant, $\theta_{\rm D}$ is the characteristic Debye temperature, and $n$ is the number of atoms in the formula unit.
The second term in Eq.~\eqref{Eq5} gives the Einstein contribution to $C_{\rm ph}$ that has the form
\begin{equation}
C_{\rm E}(\theta_{\rm E}, T) = 3nR\left(\frac{\theta_{\rm E}}{T}\right)^2 
\frac{e^{\theta_{\rm E}/T}}{[e^{\theta_{\rm E}/T}-1]^{2}}.
\label{Eq7} 
\end{equation}
Here, $\theta_{\rm E}$ is the characteristic Einstein temperature. The coefficients $f_{\rm D}$, $g_1$, $g_2$, and $g_3$ represent the fraction of atoms that contribute to their respective parts. These values are taken in such a way that their sum should be equal to 1 and are conditioned to satisfy the Dulong-Petit value $\sim 3nR$ at high temperatures. The high-$T$ fit to the $C_{\rm p}(T)$ data was then extrapolated down to low temperatures and subtracted from $C_{\rm p}(T)$. The obtained $C_{\rm mag}/T$ is plotted as a function of temperature in the main panel of Fig.~\ref{Fig5}(b) and the corresponding magnetic entropy is calculated to be $S_{\rm{mag}}(T) = \int_{\rm 0\,K}^{T}\frac{C_{\rm {mag}}(T')}{T'}dT' \simeq 11.6$~J~mol$^{-1}$~K$^{-1}$ at 12~K. This value corresponds to the expected magnetic entropy for spin-$\frac{3}{2}$: $S_{\rm mag}=R\ln 4=11.5$~J~mol$^{-1}$~K$^{-1}$. Unlike the conventional magnets which release the entire entropy near the transition temperature, LCPP releases only $\sim 40$\% of the total entropy at $T_{\rm C}$ and the remaining entropy is released only above 11~K, suggesting that $T_{\rm C}$ is partially suppressed as a result of low-dimensionality or magnetic frustration~\cite{Somesh104422}.

The inset of Fig.~\ref{Fig5}(a) presents the $C_{\rm p}(T)$ data measured in different applied fields. The influences of magnetic field is clearly reflected in the data. The zero-field peak broadens and shifts toward high temperatures with increasing field, which is usual for ferromagnets. At low temperatures, $C_{\rm mag}(T)$ in zero field could be well described using power-law ($C_{\rm mag} \propto T^{\alpha}$) behavior [inset of Fig.~\ref{Eq5}(b)] with an exponent $\alpha \sim 1.5$ that corresponds to FM spin-wave excitations~\cite{Gopal2012}.

\subsection{Magnetocaloric Effect}
\begin{figure}
	\begin{center}
		\vspace{10pt}
		\includegraphics[width=\columnwidth]{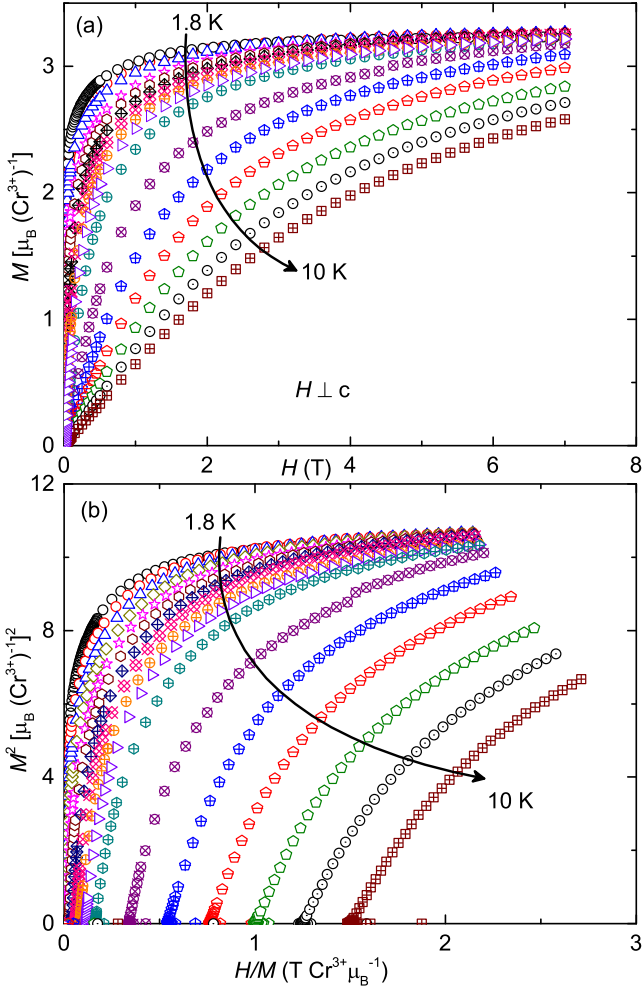} 
		\caption{(a) Isothermal magnetization ($M$ vs $H$) curves for $H\perp c$ and (b) their corresponding Arrott plots ($M^{2}$ vs $(H/M)$) for LCPP at different temperatures around $T_{\rm C}$.}
		\label{Fig6}
	\end{center}
\end{figure}
\begin{figure}
	\centering
	\includegraphics[width=\columnwidth]{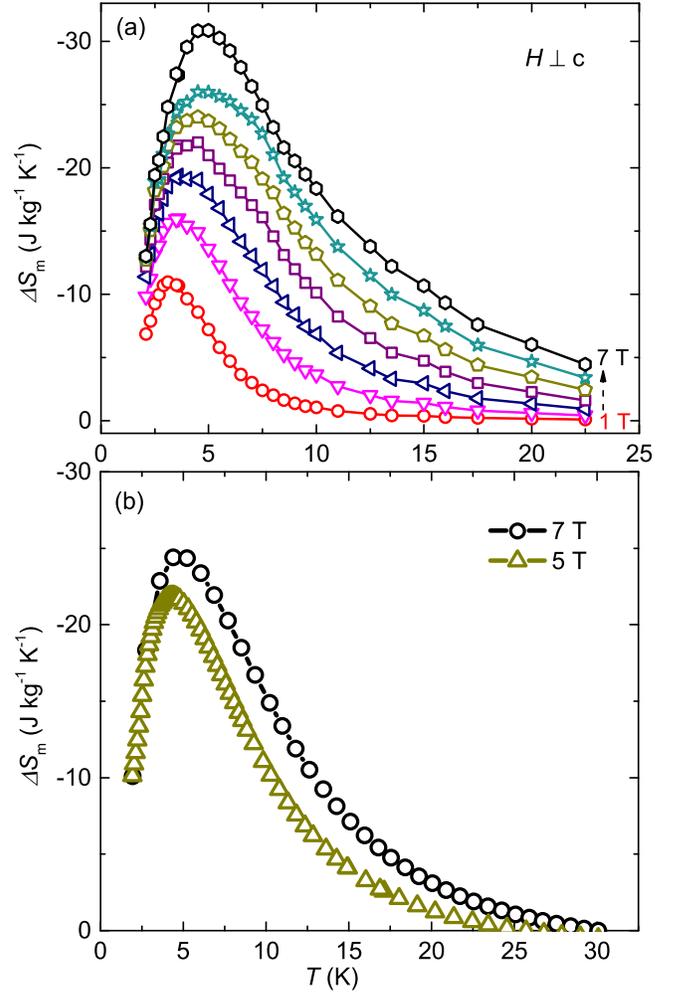}
	\caption{(a) Isothermal entropy change ($\Delta$S$_{\rm m}$) vs $T$ plotted upon sweeping the field to 0 T starting from different fields from 1~T to $7$~T, calculated using $M$ vs $H$ data of single crystals for $H \perp c$ and employing Eq.~\eqref{Sm_magnetization}. (b) $\Delta S_{\rm m}$ vs $T$ plot for $5$~T and $7$~T, calculated using $C_{\rm p}(T)$ of polycrystalline sample in Eq.~\eqref{Sm_heatcapacity}.}
	\label{Fig7}
\end{figure}
Magnetocaloric effect (MCE) is an intrinsic property of magnetic materials. Magnetic cooling is achieved by first applying magnetic field to the material isothermally and then removing the field adiabatically. Therefore, MCE is generally quantified by the isothermal entropy change ($\Delta S_{\rm m}$) and adiabatic temperature change ($\Delta T_{\rm ad}$) with respect to the change in applied field. The $\Delta S_{\rm m}$ can be calculated from either magnetization isotherms ($M$ vs $H$) or heat capacity data measured in zero and non-zero magnetic fields. Figure~\ref{Fig6}(a) displays the magnetic isotherms measured in close temperature steps around $T_{\rm C}$ for $H \perp c$. The first method utilizes Maxwell's thermodynamic relation, $(\partial S/\partial H)_T = (\partial M/\partial T)_H$, and $\Delta S_{\rm m}$ can be estimated using the $M$ vs $H$ data as~\cite{tishin2016}
\begin{equation}
\Delta S_{\rm m} (H, T) =\int_{H_{\rm i}}^{H_{\rm f}}\dfrac{dM}{dT}dH.
\label{Sm_magnetization}
\end{equation}
Figure~\ref{Fig7}(a) presents the plot of $\Delta S_{\rm m}$ as a function of temperature ($T$) in different values of $\Delta H = H_{\rm f} - H_{\rm i} $. $\Delta S_{\rm m}$ vs $T$ exhibits a maximum entropy change around 4.6~K, with a highest value of $\Delta S_{\rm m}\simeq -31$~J~kg$^{-1}$~K$^{-1}$ for the 7~T field change. As the magnetic anisotropy is negligibly small, no significant difference in $\Delta S_{\rm m}$ is expected for $H \perp c$ and $H \parallel c$~\cite{Balli232402}.

Further, to cross check the large value of $\Delta S_{\rm m}$, we have also estimated $\Delta S_{\rm m}$ from heat capacity data measured in zero field, 5~T, and 7~T. First, we calculated the total entropy at a given field as
\begin{equation}
	S(T)_H = \int_{T_{i}}^{T_{f}} \frac{ C_{\rm p}(T)_{\rm H}}{T} dT,
	\label{Sm_heatcapacity}
\end{equation}
where $C_p(T)_H$ is the heat capacity at a particular field $H$ and $T_{i}$ and $T_{f}$ are the initial and final temperatures, respectively. 
We calculated $\Delta S_{\rm m}$ by taking the difference of total entropy at non-zero and zero fields as $\Delta S_{\rm m}(T)_{\Delta \rm H} = [S(T)_{\rm H} - S(T)_{\rm 0} ]_T$. Here, $S(T)_{\rm H}$ and $S(T)_{\rm 0}$ are the total entropy in the presence of $H$ and in zero field, respectively. Figure~\ref{Fig7}(b) presents the estimated $\Delta S_{\rm m}$ as a function of temperature in 5~T and 7~T magnetic fields. The overall shape and peak position of the $\Delta S_{\rm m}$ curves are identical with the curves [Fig.~\ref{Fig7}(a)] obtained from the magnetic isotherms but with a slight reduction in magnitude. This difference in magnitude at the peak position could be related to the polycrystalline sample used for heat capacity and single crystals for magnetic measurements~\cite{Zhu144425}. 

\begin{figure}
	\centering
	\includegraphics[width=\columnwidth]{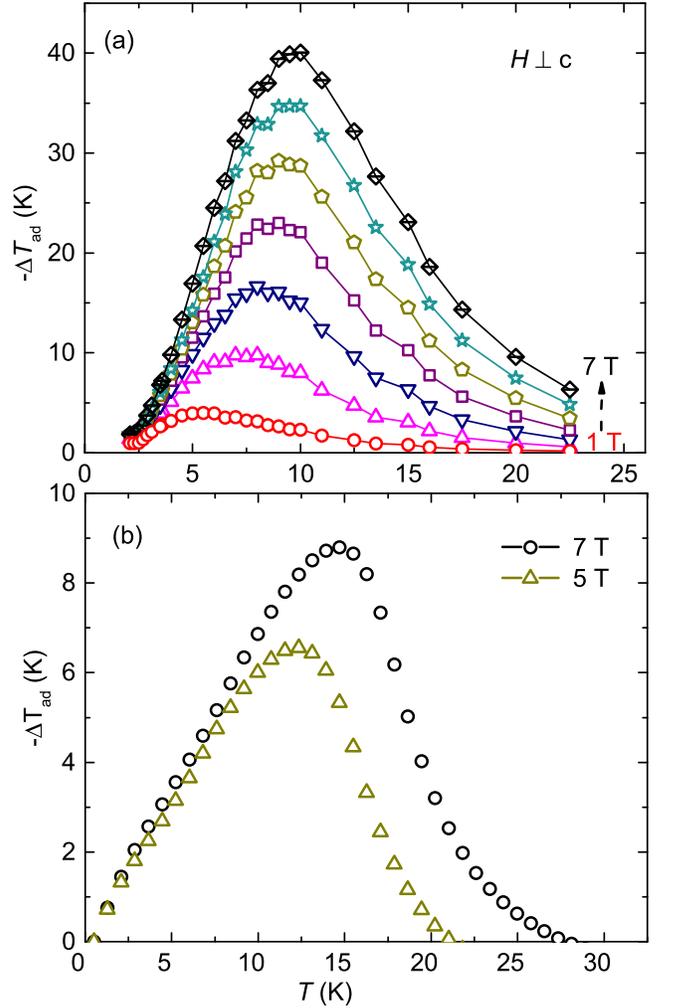}
	\caption{(a) $\Delta T_{\rm ad}$ vs $T$ plotted for different field changes of $\Delta H = 1$~T to $7$~T calculated using Eq.~\eqref{overestimated}. (b) $\Delta T_{\rm ad}$ vs $T$ plotted for $\Delta H = 5$~T and $7$~T calculated using Eq.~\eqref{proper}.}
	\label{Fig8}
\end{figure}
Similarly, the adiabatic temperature change $\Delta T_{\rm ad}$ can be estimated from either the combination of zero-field heat capacity and the magnetic entropy change obtained from magnetic isotherms or from the heat capacity alone measured in different magnetic fields. Using the heat capacity in zero field and magnetization isotherm data, the estimation of $\Delta T_{\rm ad}$ can be done as~\cite{Islam134433}
\begin{equation}
	\Delta T_{\rm ad} =\int_{H_{\rm i}}^{H_{\rm f}}\dfrac{T}{C_{\rm p}}\dfrac{dM}{dT}dH.
	\label{overestimated}
\end{equation}
The dependence of $\Delta T_{\rm ad}$ on $T$ for different magnetic fields is shown in Fig.~\ref{Fig8}(a). The maximum value of $\Delta T_{\rm ad}$ is obtained to be $\sim 40$~K for $\Delta H = 7$~T. However, as explained in Ref.~\cite{Pecharsky565} the above expression overestimates $\Delta T_{\rm ad}$ since $T/C_p$ is not constant over the range of applied fields as it was assumed. It is evident from the inset of Fig.~\ref{Fig5}(a) that $C_{\rm p}$ at low temperatures is changing drastically as we apply magnetic field and this change should be taken into account while calculating the entropy for that particular field. Therefore, we tried to estimate $\Delta T_{\rm ad}$ by taking the difference in temperatures corresponding to two different fields with constant (same) entropy as~\cite{Pecharsky565}
\begin{equation}
	\Delta T_{\rm ad}(T)_{\Delta H} = [T(S)_{H_f} - T(S)_{H_i}].
	\label{proper}
\end{equation}
$\Delta T_{\rm ad}$ vs $T$ for $\Delta H = 5$~T and $7$~T calculated by this method is shown in Fig.~\ref{Fig8}(b). The maximum value of $\Delta T_{\rm ad}$ at 7~T is around $\sim 9$~K which is significantly smaller than the value obtained using the former method [Eq.~\eqref{overestimated}]. A similar difference has been reported earlier for ErAl$_{2}$~\cite{Pecharsky565}. The latter method is considered to be more reliable. It is expected to provide accurate results as the effect of magnetic field on $C_{\rm p}$ is accounted for.

\begin{figure}
	\centering
	\includegraphics[width=\columnwidth]{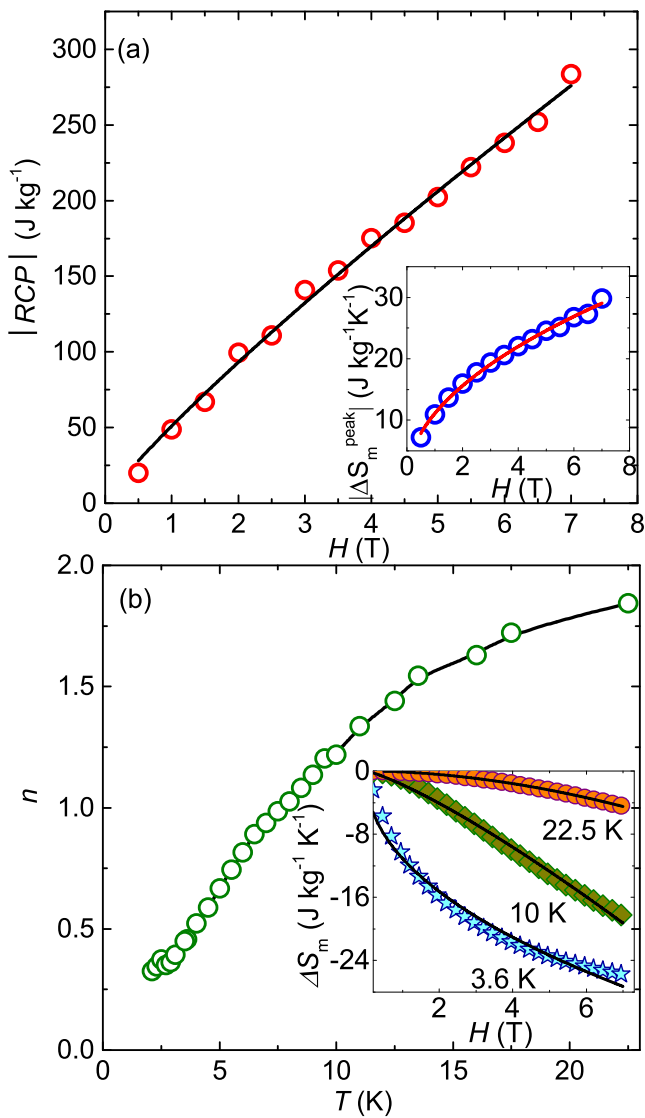}
	\caption{(a) Relative cooling power ($RCP$ = $\Delta S^{\rm peak}_m \times \delta T_{FWHM}$) as a function of magnetic field and the inset shows the value of entropy change at the peak position $\Delta S ^{\rm peak}_m$. Solid lines are the fits as described in the text. (b) The exponent $n$ plotted as a function of temperature which is obtained from fitting power law to $\Delta S_m$ vs field isotherms. Inset shows the $\Delta$S$_m$ isotherms for temperatures near and well above $T_{\rm C}$. }
	\label{Fig9}
\end{figure}
Note that the large values of $\Delta S_{\rm m}$ and $\Delta T_{\rm ad}$ are not sufficient to characterize the potential of a material for the magnetic refrigeration applications. Another important parameter is the relative cooling power ($RCP$) which is a measure of the amount of heat transferred between the cold and hot reservoirs in a refrigeration cycle. Mathematically, it can be expressed as
\begin{equation}
	{RCP} = \int_{T_{\rm {cold}}}^{T_{\rm hot}} \Delta S_{\rm m}(T,H)~dT,
\end{equation}
where, $T_{\rm cold}$ and $T_{\rm hot}$ correspond to temperatures of cold and hot reservoirs, respectively. The formula for $RCP$ can be approximated as
\begin{equation}
	|{ RCP}|_{\rm approx} = \Delta S_{\rm m}^{\rm peak} \times \delta T_{\rm FWHM},
\end{equation}
where $\Delta S_m^{\rm peak}$ and $ \delta T_{\rm FWHM}$ are the maximum value of entropy change (or the peak value) and full width at half maximum of the $\Delta S_m$ curve, respectively. $RCP$ as a function of $H$ calculated using the $\Delta S_m$ data from Fig.~\ref{Fig7}(a) is plotted in Fig.~\ref{Fig9}(a). The maximum value of $RCP$ is calculated to about $\sim 284$~J~kg$^{-1}$ at 7~T.

The application of a MCE material is also decided by the nature of its magnetic phase transition. In materials with a first-order phase transition, though the peak height of the $\Delta S_{\rm m}$ and $\Delta T_{\rm ad}$ vs $T$ curves is large but the curve width is not very broad, which limits the relevance of these materials in a cyclic operation. The second problem with the first-order transitions is the energy loss due to magnetic and thermal hysteresis~\cite{Franco305}. Further, relatively large magnetic fields are required to perturb the first-order magneto-structural transitions and induce large MCE which is another drawback of these materials. On the other hand, materials with second-order phase transition do not show very large peaks, but their $RCP$ values are large due to the increased curve width and the absence of thermal hysteresis, both effects making them promising for practical applications. In order to analyze the nature of the phase transition, we construct the Arrott plot~\cite{Nath224513} in Fig.~\ref{Fig6}(b) by using the isothermal magnetization data presented in Fig.~\ref{Fig6}(a). Clearly, the slope of $M^2$ vs $H/M$ curves is positive in the entire measured temperature range, well below and above $T_{\rm C}$. According to the Banerjee criterion~\cite{Banerjee16}, positive slope implies the second-order phase transition. This confirms the continuous second-order nature of the PM to FM phase transition in LCPP.

Furthermore, MCE is also utilized to characterize the nature of a phase transition~\cite{Islam134433,Singh6981,Singh033902}. According to the scaling hypothesis~\cite{Franco305}, the $\Delta S_m (T)$ curves for different values of $\Delta H$ should collapse on a single universal curve when the $\Delta S_m (T)$ is normalized to its peak value $\Delta S_m^{\rm peak}$. However, due to the low transition temperature, the universal curve construction is implausible with the present data. Therefore, we performed only the power-law analysis of $\Delta S_m$ and $RCP$. In Fig.~\ref{Fig9}(a) main panel and inset, we fitted the $RCP$ and $\Delta S_m^{\rm peak}$ data by power laws of the form $RCP \propto H^{N}$ and $|\Delta S_m^{\rm peak}| \propto H^n$, respectively. The exponents $N$ and $n$, which are related to the critical exponents ($\beta$, $\gamma$, and $\delta$), are estimated to be $N \simeq 0.7$ and $n \simeq 0.5$. In order to perceive the temperature dependence of $n$, we fitted the field-dependent isothermal magnetic entropy change
$\Delta S_m(H)$ at various temperatures across the transition using the power law $\Delta S_m \propto H^n$ [see inset of Fig.~\ref{Fig9}(b)]~\cite{Singh6981}. The obtained $n$ vs $T$ data are plotted in Fig.~\ref{Fig9}(b) and provide information concerning the nature of the transition. For instance, for a second-order magnetic transition the exponent should have the value $n \simeq 2$ in the paramagnetic region ($T >> T_{\rm C}$) and $n(T)$ typically exhibits a minimum near $T_{\rm C}$~\cite{Singh6981}. Indeed, our $n(T)$ demonstrates the expected behavior, further confirming the second-order magnetic transition in LCPP.
\begin{table*}
	\caption{Comparison of the adiabatic temperature change ($\Delta T_{\rm ad}$), maximum entropy change ($\Delta S_m^{\rm peak}$), and relative cooling power ($RCP$) of LCPP with some known magnets having low transition temperatures ($T_{\rm C}$ or $T_{\rm N}$) and large MCE in a field change of $\Delta H = 5$ to 8~T. More compounds with the similar behaviour are listed in Refs.~\cite{Midya094422,Zhu144425,Zheng1462}.}
	\begin{ruledtabular}
	\begin{tabular}{ccccccccccccc}
		System & $T_{\rm C}/T_{\rm N}$ & $|\Delta T_{\rm ad}|$ & $|\Delta S_{\rm m}^{\rm peak}|$ & $RCP$ & $\Delta H$ & Ref. \\
		& (K)   & (K)                 & (J~kg$^{-1}$~K$^{-1}$)            & (J~kg$^{-1}$) & (T) & \\
		\hline
		LCPP & 2.6 & 9 & 31 & 284 & 7 & This work \\
		HoMnO$_3$ & 5 & 6.5 & 13.1 & 320 & 7 & \cite{Midya142514}\\
		ErMn$_2$Si$_2$ & 4.5 & 12.9 &	25.2 & 365 & 5 & \cite{Lingwei152403}\\	
		EuTi$_{0.9}$V$_{0.3}$O$_3$ & 4.5 & 17.4 & 41.4 & 577 & 7 & \cite{Roy064412}\\
		GdCrTiO$_5$ & 0.9 & 15.5 & 36 & -- & 7 &  \cite{Das104420}\\	
		EdDy$_2$O$_4$ &	5 & 16 & 25 & 415 & 8 & \cite{Midya132415}\\
		EuHo$_2$O$_4$ &	5 &	12.7 & 30 &	540	& 8 & \cite{Midya132415}\\
		Mn$_{32}$ &0.32& 6.7 & 18.2 &-- & 7 & \cite{Evangelisti104414}\\
		HoB$_{2}$ &15& 12 & 40.1 &- & 5 & \cite{Castro35}\\
		EuTiO$_{3}$ & 5.6 & 21 & 49 & 500 & 7 & \cite{Midya094422}\\
	\end{tabular}
\end{ruledtabular}
	\label{TableIII}
\end{table*}

In Table~\ref{TableIII}, we compare the main parameters of LCPP with those of well-studied magnets having low transition temperatures and large MCE. Though the $\Delta S_{\rm m}^{\rm peak}$ value of LCPP is comparable to the values for most of the potential low-temperature magnetic refrigerant materials, the width of $\Delta S_{\rm m}$ vs $T$ curves is not very broad. Due to which LCPP has a slightly reduced value of $RCP$ compared to others. Nevertheless, the obtained value of $RCP \simeq 284$~J~kg$^{-1}$ is still significantly large and LCPP may have strong prerequisites for cryogenic applications in sub-Kelvin temperatures~\cite{Gschneidner1479,Martinez4301}.

\subsection{Electron Spin Resonance}
\subsubsection{Frequency Dependence at $T = 1.8$~K}
\begin{figure}
	\includegraphics[width=\columnwidth]{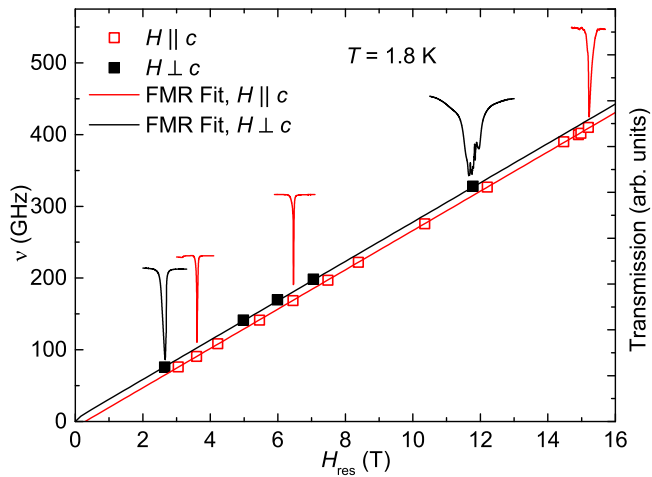}
	\caption{Left vertical scale: Frequency as a function of resonant field measured for both orientations at $T = 1.8$~K. The hollow red and solid black squares represent the measured resonance fields for orientations $H \parallel c$ and $H \perp c$, respectively. Right vertical scale: Selected HF-ESR spectra.}
	\label{Fig10}
\end{figure}

The ferromagnetic resonance (FMR) measurements on LCPP were performed at $T = 1.8$~K ($< T_{\rm C}$) for the two orientations of the applied magnetic field, $H \parallel c$ and $H \perp c$. The frequency ($\nu$) vs resonance field ($H_{\rm res}$) dependence of the FMR signal is shown in Fig.~\ref{Fig10} together with few selected spectra (right axis).  The spectra measured at fields perpendicular to the $c$-axis are relatively broader and more distorted when compared to the $H \parallel c$ measurements. The $\nu$ vs $H_{\rm res}$ data were fitted with a spin-wave model for FMR, $\nu = h^{-1}g_{\parallel}\mu_{\rm B} (H_{\rm res} - H_{\rm a})$ for $H \parallel c$ orientation and $\nu = h^{-1}g_{\perp}\mu_{\rm B} [H_{\rm res}(H_{\rm res} + H_{\rm a})]^{1/2}$ for $H \perp c$ orientation to obtain the $g$-factors and the anisotropy field $H_{\rm a}$. It was found that the $g$-factors in both orientations are somewhat different amounting to $g_{\perp} = 1.968 \pm 0.003$ and $g_{\parallel} = 1.958 \pm 0.003$. Such a slightly anisotropic $g$-tensor is typical for a Cr$^{3+}$ ion ($3d^3$, $S = 3/2$, $L = 3$) in a distorted octahedral ligand coordination~\cite{Abragam2012}.
Further, the anisotropy field $H_{\rm a}$ was obtained as $\sim 2864$~Oe from the fit. Generally, $H_{\rm a}$ consists of two contributions~\cite{Farle1998}: 
\begin{equation}
	H_{\rm a} = 4{\pi}M-2K/M.
	\label{eq:Ha}
\end{equation}
The first term accounts for the shape anisotropy and the second term is the intrinsic magnetocrystalline anisotropy. In Eq.~\eqref{eq:Ha}, the first term assumes the shape anisotropy of a thin plate with $N=1$.

Using the saturation magnetization value, $M_{\rm s} = g_\parallel S = 2.94~\mu_{\rm B}/{\rm Cr}^{3+}$, the magnetocrystalline anisotropy constant $K = -7.42 \times 10^{4}$~erg~cm$^{-3}$ was obtained. The negative sign of $K$ implies that LCPP is an easy-plane ferromagnet with the hard magnetic axis normal to the $ab$-plane. We note that no other resonance excitations could be found in a frequency range up to 975~GHz (4~meV).

\begin{figure}
	\centering
	\includegraphics[width=\columnwidth]{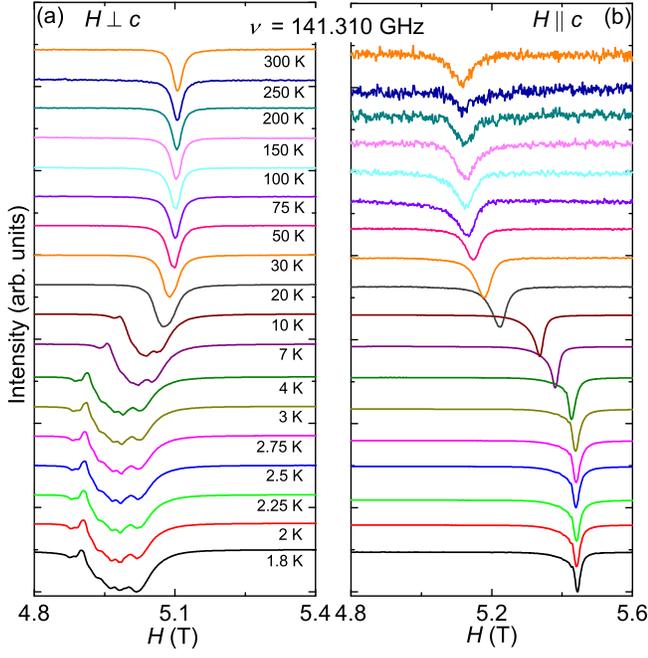}
	\caption{Temperature dependence of the HF-ESR spectra at an excitation frequency of 141.310~GHz for (a) $H \perp c$ and (b) $H \parallel c$. Line shape distortions at $T < 20$~K in the $H \perp c $ field geometry are instrumental artifacts arising due to the strong magnetization of the sample.}
	\label{Fig11}
\end{figure}
\begin{figure}[h]
	\centering
	\includegraphics[width=\columnwidth]{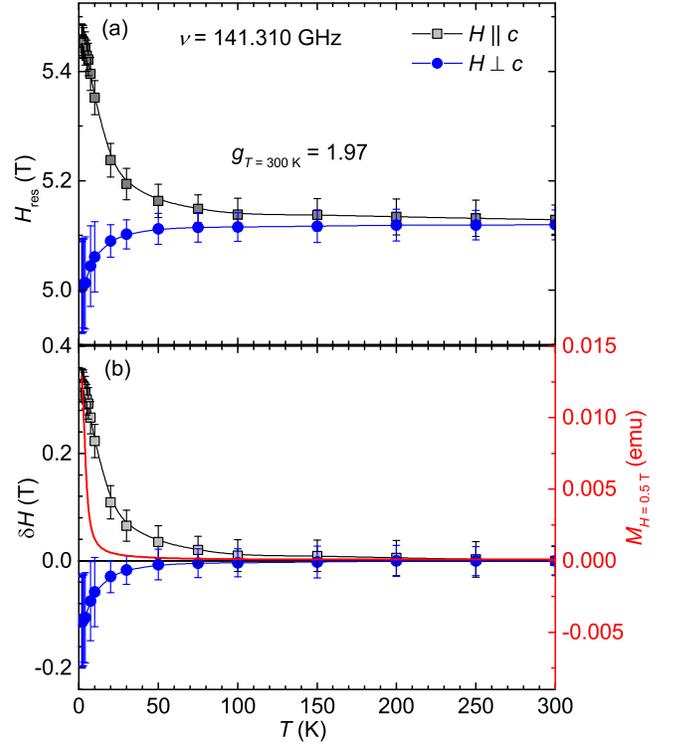}
	\caption{Temperature dependence of (a) the resonance field $H_{\rm res}$ and (b) the resonance shift $\delta H = H_{\rm res}(T) - H_{\rm res}(300~{\rm K})$ (left $y$-axis) and the magnetization $M$ in an applied field of $H = 0.5$~T (right $y$-axis), for both $H \parallel c$ and $H \perp c$.}
	\label{Fig12} 
\end{figure}
\subsubsection{Temperature Dependence}
HF-ESR spectra of LCPP at various temperatures were measured at a fixed excitation frequency of $\nu = 141.310$~GHz for both field directions [see Figs.~\ref{Fig11}(a) and (b)]. The resonance fields $H_{\rm res}$ were obtained from the absorption minima of each spectrum and are plotted against temperature in Fig.~\ref{Fig12}(a). Interestingly, the $H_{\rm res}$ vs $T$ dependence for  both orientations does not converge to a paramagnetic line immediately at $T_{\rm C} = 2.8$~K. Only above 10~K, the resonance fields for $H \parallel c$ and $H \perp c$ orientations rapidly start to decrease and increase, respectively, toward the expected paramagnetic position and almost merge around 100~K at a field corresponding to the  $g$-factor, $g = 1.97$. The lineshape distortions in measurements at low temperatures for $H \perp c$ are accounted for in the enlarged error bars.

The shift of the resonance position $\delta H(T)$ from the paramagnetic one is shown in Fig.~\ref{Fig12}(b) (left $y$-axis). $\delta H(T)$ is positive and larger when the external field is parallel to the magnetic hard axis, as compared to the smaller negative shift for the in-plane field geometry, as expected for an easy-plane ferromagnet. Such a shift cannot be ascribed entirely to the shape anisotropy, which should play a role also in the paramagnetic state if the sample's magnetization $M$ is large. The $M(T)$ curve plotted in Fig.~\ref{Fig12}(b) (right $y$-axis) for comparison decreases right above $T_{\rm C}$ more rapidly than the $\delta H(T)$ dependence. Therefore, the line shift observed for both orientations well above the Curie temperature may be indicative of short-range FM spin correlations on the fast ESR time scale, typical for low-dimensional magnets such as, e.g., the quasi-two-dimensional van der Waals compound Cr$_2$Ge$_2$Te$_6$~\cite{Zeisner2019}. This is also the case for LCPP owing to its layered crystal structure.

\subsection{Microscopic Analysis}
Our DFT calculations return nearest-neighbor exchange coupling $J/k_{\rm B} \simeq -0.5$~K within the kagome planes. This value is somewhat dependent on the choice of the DFT+$U$ parameters, but the negative sign is robust and suggests ferromagnetic nature of the kagome network in LCPP. The origin of this ferromagnetic coupling deserves some attention, as the sibling Fe$^{3+}$ compound is clearly antiferromagnetic~\cite{Kermarrec157202}. Superexchange theory stipulates that antiferromagnetic couplings are mediated by hoppings between half-filled orbitals, whereas ferromagnetic couplings arise from hoppings between the half-filled and empty orbitals. In LCPP with the $3d^3$ Cr$^{3+}$ magnetic ion, these orbitals have the $t_{2g}$ and $e_g^{\sigma}$ characters, respectively. Trigonal symmetry of the crystal structure further splits the $t_{2g}$ levels into $a_{1g}$ and $e_g^{\pi}$. 
\begin{figure}[h]
	\includegraphics{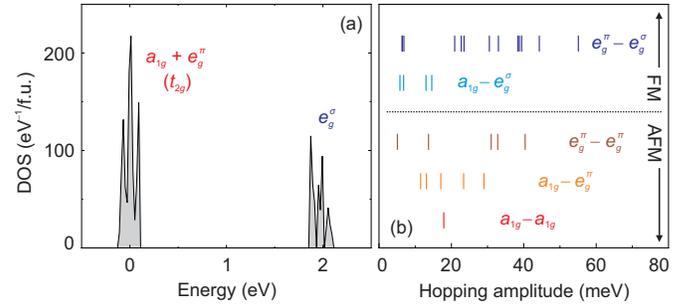}
	\caption{\label{Fig13}
		(a) PBE density of states for LCPP with the Fermi level placed at zero energy. (b) Nearest-neighbor Cr--Cr hopping amplitudes within the kagome plane.}
\end{figure}

\begin{figure}[h]
	\includegraphics{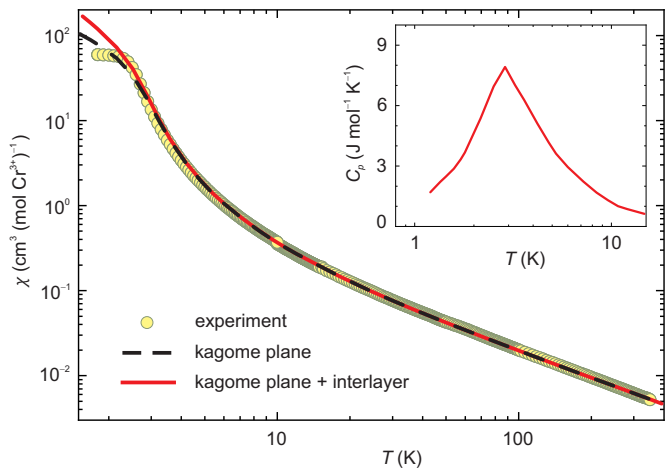}
	\caption{\label{Fig14}
		Magnetic susceptibility of LCPP measured in the applied field of 0.01\,T upon field cooling and its fit using the model of ferromagnetic kagome planes (dashed line) and coupled ferromagnetic kagome planes ($J_{\perp}/J=0.01$, solid line). The inset shows calculated specific heat for the coupled kagome planes, with $T_C \simeq 2.8$\,K.}
\end{figure}
DFT band structure of LCPP calculated on the PBE level features narrow $t_{2g}$ ($a_{1g}+e_g^{\pi}$) bands around the Fermi level and almost equally narrow $e_g^{\sigma}$ bands centered at around 2.0\,eV (Fig.~\ref{Fig13}). This band structure is metallic because neither magnetism nor correlation effects have been taken into account. The small band width of 0.25\,eV for the $t_{2g}$ bands indicates that antiferromagnetic contribution to the exchange couplings should be minor. The $t_{2g}-e_g^{\sigma}$ hoppings are small too, but somewhat larger than the $t_{2g}-t_{2g}$ hoppings, as shown in Fig.~\ref{Fig13}(b). Therefore, ferromagnetic contribution to the exchange becomes predominant, and the overall coupling is ferromagnetic. Magnetic susceptibility of LCPP is well described by the model of nearest-neighbor ferromagnetic kagome planes with $J/k_{\rm B} =-1.2$\,K ($g=1.995$). A minute interlayer coupling $J_{\perp}/J=0.01$ improves the fit below 3.5\,K and leads to the Curie temperature $T_C=2.8$\,K in a perfect agreement with the experimental $T_{\rm C}\simeq 2.6$\,K (Fig.~\ref{Fig14}). Our DFT calculations corroborate this result and reveal a weakly ferromagnetic $J_{\perp}$ with $J_{\perp}/J\simeq 0.02$. 

\section{Summary}
We synthesized single crystals of LCPP and confirmed trigonal $P\bar{3}c1$ symmetry of this compound with the lattice constants $a = b = 9.668(3)$~\AA\ and $c = 13.610(6)$~\AA\ at room temperature. Green-colored LCPP is a rare example of an insulating kagome ferromagnet. Ferromagnetic order below $T_{C} \simeq 2.6$~K is driven by the in-plane FM coupling $J/k_{\rm B} \simeq -1.2$~K supplied with a minute inter-plane coupling $J_{\perp}/J=0.02$, which is also FM in nature. The incomplete release of the magnetic entropy at $T_{\rm C}$ and the increased width of the ESR line above $T_{\rm C}$ both suggest quasi-2D magnetic behavior caused by the strong spatial anisotropy of FM couplings. The overall magnetic behavior of LCPP has striking resemblance with that of other insulating kagome ferromagnets, such as $\alpha$-MgCu$_3$(OD)$_6$Cl$_2$ and Cu[1,3-bdc]~\cite{Boldrin220408,Chisnell214403}. Magnetization and ESR measurements on single crystals indicate a weak easy-plane anisotropy. The critical scaling of magnetization suggests a non-mean-field type second-order nature of the phase transition at $T_{\rm C}$. The low $T_{\rm C}$ combined with the large values of $\Delta S_{\rm m}$, $\Delta T_{\rm {ad}}$, and $RCP$ render LCPP a promising magnetocaloric material for low-temperature applications.

\acknowledgments
AM, SK, VS, and RN would like to acknowledge SERB, India for financial support bearing sanction Grant No.~CRG/2019/000960.

%

\end{document}